\documentclass[epj]{webofc}
\usepackage[utf8]{inputenc}
\usepackage[varg]{txfonts}
\usepackage{booktabs}

\usepackage{xcolor}
\definecolor{darkred}{rgb}{0.4,0.0,0.0}
\definecolor{darkgreen}{rgb}{0.0,0.4,0.0}
\definecolor{darkblue}{rgb}{0.0,0.0,0.4}
\usepackage[bookmarks,linktocpage,colorlinks,linkcolor=darkred,urlcolor=darkblue,citecolor=darkgreen]{hyperref}
\usepackage{changepage}

\wocname{EPJ Web of Conferences}
\woctitle{Lattice2017}

\newcommand{\cstar}{C$^\star$\,}
\newcommand{\psibar}{\bar{\psi}}
\newcommand{\csw}{\mathrm{c}_\mathrm{sw}}

\begin{document}
\title{Simulations of QCD and QED with \cstar boundary conditions}

\author{
\firstname{Martin}~\lastname{Hansen}\inst{1}\fnsep\thanks{Speaker, \email{hansen@cp3.sdu.dk}}
\and
\firstname{Biagio}~\lastname{Lucini}\inst{2}
\and
\firstname{Agostino}~\lastname{Patella}\inst{3,4}
\and
\firstname{Nazario}~\lastname{Tantalo}\inst{5}
}

\institute{
\footnotesize
CP3-Origins, University of Southern Denmark, Campusvej 55, DK-5230 Odense M, Denmark
\and
College of Science, Swansea University, Singleton Park, Swansea, SA2 8PP, UK
\and
Theoretical Physics Department, CERN, Geneva, Switzerland
\and
Centre for Mathematical Sciences, Plymouth University, Plymouth, PL4 8AA, UK
\and
Dipartimento di Fisica and INFN, Università di Roma ``Tor Vergata'', Via della Ricerca Scientifica 1, I-00133 Rome, Italy
}

\abstract{
We present exploratory results from dynamical simulations of QCD in isolation, as well as QCD coupled to QED, with \cstar boundary conditions. In finite volume, the use of \cstar boundary conditions allows for a gauge invariant and local formulation of QED without zero modes. In particular we show that the simulations reproduce known results and that masses of charged mesons can be extracted in a completely gauge invariant way. For the simulations we use a modified version of the HiRep code. The primary features of the simulation code are presented and we discuss some details regarding the implementation of \cstar boundary conditions and the simulated lattice action.
\\[1mm]
{\footnotesize Preprint: CP$^3$-Origins-2017-046 DNRF90, CERN-TH-2017-214} \\
}

\maketitle
\vspace{-8mm}

\section{Introduction}\label{intro}\vspace{-1mm}
Calculating electromagnetic corrections to hadronic observables via lattice simulations requires a consistent formulation of QCD+QED in finite volume. Such a formulation, based on \cstar boundary conditions, was proposed and thoroughly discussed in \citep{Lucini:2015hfa}. Here we present the first exploratory simulations showing that the proposed setup works in practice. For the simulations we use a modified version of the HiRep code, as discussed in section \ref{code}, and in section \ref{studies} we present our exploratory results. In particular, we show that masses of charged hadrons can be extracted in a gauge invariant way, with a signal-to-noise ratio equivalent to that of neutral hadrons.
\vspace{-1mm}
\section{Simulation code}\label{code}\vspace{-1mm}
In this section we present a modified version of the HiRep code \cite{DelDebbio:2008zf} used for simulating QCD and QED with \cstar boundary conditions. In particular we outline the primary features of the code and discuss the approach used to implement the boundary conditions.
\vspace{-1mm}
\subsection{Overview}
The HiRep code was developed as a flexible code for studying BSM models and for this reason it has native support for different gauge groups and higher dimensional fermion representations. Especially the latter feature was one of the main reasons for choosing to modify this code, as discussed later on. The modified code presented here can simulate QCD, QED and QCD+QED with either periodic or \cstar boundary conditions in space. The main features of the code include: \\[-3mm]
\begin{adjustwidth}{5mm}{}
\begin{itemize}
 \item Wilson fermions with $O(a)$ improvement
 \item L\"uscher-Weisz gauge action for SU(3) field
 \item Plaquette gauge action for U(1) field
 \item Compact QED with optional Fourier acceleration
 \item Support for rational approximations
 \item Hierarchical OMF integrators
 \item Several inverters, such as MINRES, BiCGstab, and CG with multishift support
 \item Gradient flow observables for both SU(3) and U(1) field
 \item Measurements of charged and neutral meson correlators
\end{itemize}
\end{adjustwidth}
~\\[-3mm]\noindent
The approach used for implementing the \cstar boundary conditions has been described in the appendix of \cite{Lucini:2015hfa}. For the discussion at hand we recall that \cstar boundary conditions correspond to performing a charge conjugation when wrapping around the torus. Denoting the matter fields by $\psi_f$ and the U(1) and SU(3) gauge potentials by $A_\mu$ and $B_\mu$, respectively, the boundary conditions read:
\begin{align}
 A_\mu(x+\hat{L}_k) &= -A_\mu(x) \\
 B_\mu(x+\hat{L}_k) &= -B_\mu^*(x) \\
 \psi_f(x+\hat{L}_k) &= C^{-1}\psibar_f^T(x) \\
 \psibar_f(x+\hat{L}_k) &= -\psi_f^T(x)C
\end{align}
Here $C$ is the charge conjugation matrix. These boundary conditions pose no additional complications for the gauge fields. In terms of the link variables, for both gauge fields, the boundary conditions correspond to complex conjugating the links when wrapping around the torus.
\begin{equation}
 U_\mu(x+\hat{L}_k) = U_\mu^*(x)
\end{equation}
In practice this constraint is implemented by having an extended lattice with a static border around the actual lattice. The relevant field variables are then copied into the static border and the boundary conditions are applied. This operation must be repeated every time the field changes.

\subsection{Fermion representation}\label{fermionrep}
While the boundary conditions for the gauge fields are trivially implemented, special care is needed for the matter fields because of the mixing between fermions and antifermions at the boundary. Due to this mixing it is advantageous to write the fermion action in terms of a new doublet $\chi$ containing both fermion fields.
\begin{equation}
 \chi = \begin{pmatrix} \psi \\ C^{-1}\psibar^T \end{pmatrix}
\end{equation}
In this notation the \cstar boundary conditions simply swap the two components of the doublet.
\begin{equation}
 \chi(x+\hat{L}_k) = \begin{pmatrix} 0 & 1 \\ 1 & 0 \end{pmatrix}\chi(x)
\end{equation}
The boundary conditions can be diagonalized via a unitary transformation in which case the basis is the two eigenstates $\psi_\pm$ of the charge conjugation operator.
\begin{equation}
\eta = \begin{pmatrix} \psi_+ \\ -i\psi_- \end{pmatrix} =  \frac{1}{\sqrt{2}}\begin{pmatrix} \psi + C^{-1}\bar{\psi}^T \\ -i(\psi - C^{-1}\bar{\psi}^T) \end{pmatrix}
\end{equation}
In this basis the boundary conditions read
\begin{equation}
 \eta(x+\hat{L}_k) = \begin{pmatrix} 1 & 0 \\ 0 & -1 \end{pmatrix}\eta(x)~,
\end{equation}
and the link variables appearing the Dirac operator are $6\times6$ real matrices constructed via the map
\begin{equation}
 U_\eta = \begin{pmatrix} \mathrm{Re}~U & -\mathrm{Im}~U \\ \mathrm{Im}~U & \mathrm{Re}~U \end{pmatrix}~.
 \label{eq:map}
\end{equation}
The relation $D[U]^T=CD[U^*]C^{-1}$ can now be used to rewrite the action.
\begin{equation}
 S_F = \psibar D[U]\psi = -\frac{1}{2}\eta^TCD[U_\eta]\eta
\end{equation}
Because the boundary conditions have been diagonalized we can evaluate the associated path integral. This results in the Pfaffian of the Dirac operator instead of the usual determinant.
\begin{equation}
 \int [D\eta]~\exp\left\{+\frac{1}{2}\eta^TCD[U_\eta]\eta\right\}
 = \mathrm{Pf}~CD[U_\eta]
\end{equation}
Using the fact that $CD[U_\eta]$ is an antisymmetric matrix, the absolute value of the Pfaffian can be related to the determinant via the identification
\begin{equation}
 \left|\mathrm{Pf}~CD[U_\eta]\right| = \sqrt{\det D[U_\eta]}~.
\end{equation}
Due to this relation, on the lattice we can use the pseudofermion method to represent the Pfaffian, but it requires the use of rational approximations in the action.
\begin{equation}
 \left|\mathrm{Pf}~CD[U_\eta]\right| = \int [D\phi][D\phi^*]~\exp\left\{-\phi^\dagger\{D[U_\eta]^\dagger D[U_\eta]\}^{-1/4}\phi\right\}
\end{equation}
Because the HiRep code already supported higher dimensional representations, defining the new representation $U_\eta$ used in the Dirac operator, amounted to writing a new function that implements the map defined in Eq. \eqref{eq:map}.

\subsection{Lattice action}
Because the quarks are fractionally charged, when using the compact formulation of QED it is necessary to rescale the elementary charge to ensure gauge covariance. We define the U(1) action as
\begin{equation}
S_G^{QED} = \beta_\mathrm{em}\sum_x\sum_{\nu<\mu}[1-\mathrm{Re}~P_{\mu\nu}(x)]~,
\end{equation}
where the bare coupling $\beta_\mathrm{em}$ now depends on the electromagnetic coupling constant $\alpha$ and the elementary charge $q_\mathrm{el}$. The latter is a tunable parameter chosen to be $q_\textrm{el}=1/6$ in our simulations.
\begin{equation}
 \beta_\mathrm{em} = \frac{1}{4\pi\alpha q_\mathrm{el}^2}
\end{equation}
The Dirac operator used for QCD+QED simulations is written in terms of the U(3) links defined by
\begin{equation}
 W_\mu(x) = U_\mu(x)V_\mu(x)^{\hat{q}}~,
\end{equation}
where $V_\mu(x)$ is the U(1) link and $U_\mu(x)$ is the SU(3) link. The integer exponent $\hat{q}=q/q_\mathrm{el}$ is the fermion charge in units of the elementary charge. With these definitions, the $O(a)$ improved Dirac operator can be written as
\begin{align}
D\phi(x) &= (4+m_0)\phi(x)
 - \frac{1}{2}\sum_\mu(1-\gamma_\mu)W_\mu(x)\phi(x+\hat{\mu}) + (1+\gamma_\mu)W_\mu^\dagger(x-\hat{\mu})\phi(x-\hat{\mu}) \\
 &+ \frac{i}{4}\sum_{\mu,\nu}\sigma_{\mu\nu}\left\{\textrm{c}_\textrm{sw}^\textrm{QCD}\hat{F}^\mathrm{QCD}_{\mu\nu}(x) + \textrm{c}_\textrm{sw}^\textrm{QED}\hat{F}^\mathrm{QED}_{\mu\nu}(x)\right\}\phi(x)~,
\end{align}
where we use the clover definition of the field tensors. For the simulations presented in the next section, we use the tree-level coefficient for the U(1) clover term given by $\textrm{c}_\textrm{sw}^\textrm{QED} = \hat{q}$.

\subsection{Interpolating operators}\label{interp}
The construction of gauge-invariant interpolating operators for charged states was discussed in \citep{Lucini:2015hfa}. In the next section we use the so-called ``Coulomb operator'' for measurements of charged meson states.
\begin{equation}
 \Psi(x) = \exp\left\{-iq\int d^3y~\partial_k A_k(x_0,\mathbf{y})\Phi(\mathbf{y}-\mathbf{x})\right\}\psi(x)
\end{equation}
The electrostatic potential $\Phi(\mathbf{x})$ must be anti-periodic and satisfy $\partial_k\partial_k\Phi(\mathbf{x}) = \delta^3(\mathbf{x})$. This operator is invariant under local gauge transformations, but transforms under global gauge transformations, and as such it defines a charged state. In the Coulomb gauge $\Psi(x)=\psi(x)$, and the operator $\Psi(x)$ is therefore the unique gauge-invariant extension of the quark field defined in the Coulomb gauge. We refer again to \citep{Lucini:2015hfa} for a definition of the discretized operator implemented in the code.

\section{Exploratory studies}\label{studies}
In this section we discuss some exploratory simulations of QCD with \cstar boundary conditions as well as dynamical QCD+QED simulations. For the QCD simulations we use parameters from the CLS collaboration to allow for a comparison with their results. For the QCD+QED simulations we start from the parameters of one of the QCD simulations, while adding the QED interactions in the Dirac operator. Introducing the QED interactions will, most notably, shift the value of the critical masses, and hence affect the mass spectrum. We study how the mass spectrum changes when adding the QED interactions, and in particular, we show that the masses of charged hadrons can be extracted in a completely gauge invariant way.

\subsection{QCD}
We have performed several QCD simulations to show that our setup with \cstar boundary conditions works in practice. For the simulations we have chosen $N_f=3$ dynamical fermions at the isospin symmetric point with the bare parameters taken from the CLS collaboration \cite{Bruno:2014jqa}. The simulated ensembles are listed in Table~\ref{tab:qcdparam} together with the bare parameters and the number of thermalized configurations used in the analysis.
\begin{table}[tbh]
\centering
\def\arraystretch{1.3}
\begin{tabular}{c|cccccc}
 Ensemble & $L^3\times T$ & $\beta$ & $\kappa$ & $\csw$ & b.c.s. & MDU \\
 \hline\hline
 A1 & $16^3\times 32$ & 3.40 & 0.13675962 & 1.986246 & PPPP & 1000 \\
 A2 & $16^3\times 32$ & 3.40 & 0.13675962 & 1.986246 & CCCP & 1500 \\
 A3 & $24^3\times 48$ & 3.40 & 0.13675962 & 1.986246 & CCCP &  693 \\
 \hline
 B1 & $16^3\times 32$ & 3.55 & 0.13700000 & 1.824865 & CCCP & 1000 \\
 B2 & $24^3\times 48$ & 3.55 & 0.13700000 & 1.824865 & CCCP & 1000 \\
\end{tabular}
\caption{Ensembles and parameters for the QCD simulations.}
\label{tab:qcdparam}
\end{table}

Except from ensemble A1 we use \cstar boundary conditions in the spatial directions, and periodic boundary conditions in time. Ensemble A1 was simulated with periodic boundary conditions in all directions to allow for a direct comparison between the two cases. For all ensembles we have measured the PCAC mass, the pion/kaon mass and decay constant and the gradient flow observable $t_0$. The results are listed in Table~\ref{tab:qcdresult}, where the PCAC mass is unrenormalized, but $O(a)$ improved using $c_A$ from \cite{Bulava:2015bxa}, and the decay constant has been renormalized using $Z_A$ from \cite{Bulava:2016ktf}.
\begin{table}[tbh]
\centering
\def\arraystretch{1.3}
\begin{tabular}{c|ccccc}
 Ensemble & $m_q$ & $m_{\pi,K}$ & $f_{\pi,K}$ & $t_0$ & $m_\pi L$\\
 \hline\hline
 A1   & 0.00868(35)  & 0.2006(82)  & 0.0486(21)  & 2.960(48)  & 3.2 \\
 A2   & 0.00877(24)  & 0.1820(47)  & 0.0562(12)  & 2.879(34)  & 2.9 \\
 A3   & 0.00901(12)  & 0.1819(16)  & 0.06022(96) & 2.901(10)  & 4.3 \\
 \hline
 B1   & 0.00665(23)  & 0.1866(65)  & 0.0305(24)  & 5.01(10)   & 3.0 \\
 B2   & 0.006805(93) & 0.1324(28)  & 0.04597(51) & 5.259(41)  & 3.2 \\
\end{tabular}
\caption{Results for the QCD simulations.}
\label{tab:qcdresult}
\end{table}

\noindent The results for the A (B) ensembles should be compared with CLS ensemble H101 (H200) in \cite{Bruno:2016plf}. Because our volumes are smaller than the corresponding CLS ensembles we have significant finite volume effects, especially on the B1 ensemble. On the two largest volumes A3 and B2 our results agree with the CLS values within 5\% in all cases. This seems reasonable given the smaller volume and smaller amount of statistics.

\subsection{QCD+QED}
For our first test simulations of dynamical QCD+QED we took the parameters for the B1 ensemble and added the QED interactions. While, in these simulations, the bare masses are still degenerate for all three quarks, in the Dirac operator we use the physical charges i.e.~we have one up-type quark with charge $q=2/3$ and two down-type quarks with $q=-1/3$. Because the charges are different, the quark masses are also renormalized differently, and for this reason it makes little sense to keep the bare masses degenerate. However, for these exploratory simulations we just wanted a simple setup for studying the two-point functions.

In Table~\ref{tab:qcdqedparam} we list the two simulated QCD+QED ensembles, with the only difference being the value of electromagnetic coupling constant. In the first ensemble we use an unphysically large value of the coupling constant $\alpha\approx 7\alpha_\mathrm{phys}$ and in the second ensemble we use the physical value.
\begin{table}[tbh]
\centering
\def\arraystretch{1.3}
\begin{tabular}{c|ccccccc}
 Ensemble & $L^3\times T$ & $\beta$ & $\alpha$ & $\kappa$ & $\csw^\mathrm{QCD}$ & b.c.s. & MDU \\
 \hline\hline
 Q1 & $16^3\times 32$ & 3.55 & 0.05 & 0.13700000 & 1.824865 & CCCP & 1000 \\
 Q2 & $16^3\times 32$ & 3.55 & 1/137 & 0.13700000 & 1.824865 & CCCP & 500 \\
\end{tabular}
\caption{Ensembles and parameters for the QCD+QED simulations.}
\label{tab:qcdqedparam}
\end{table}

\begin{figure}
 \centering
 \includegraphics[scale=0.45]{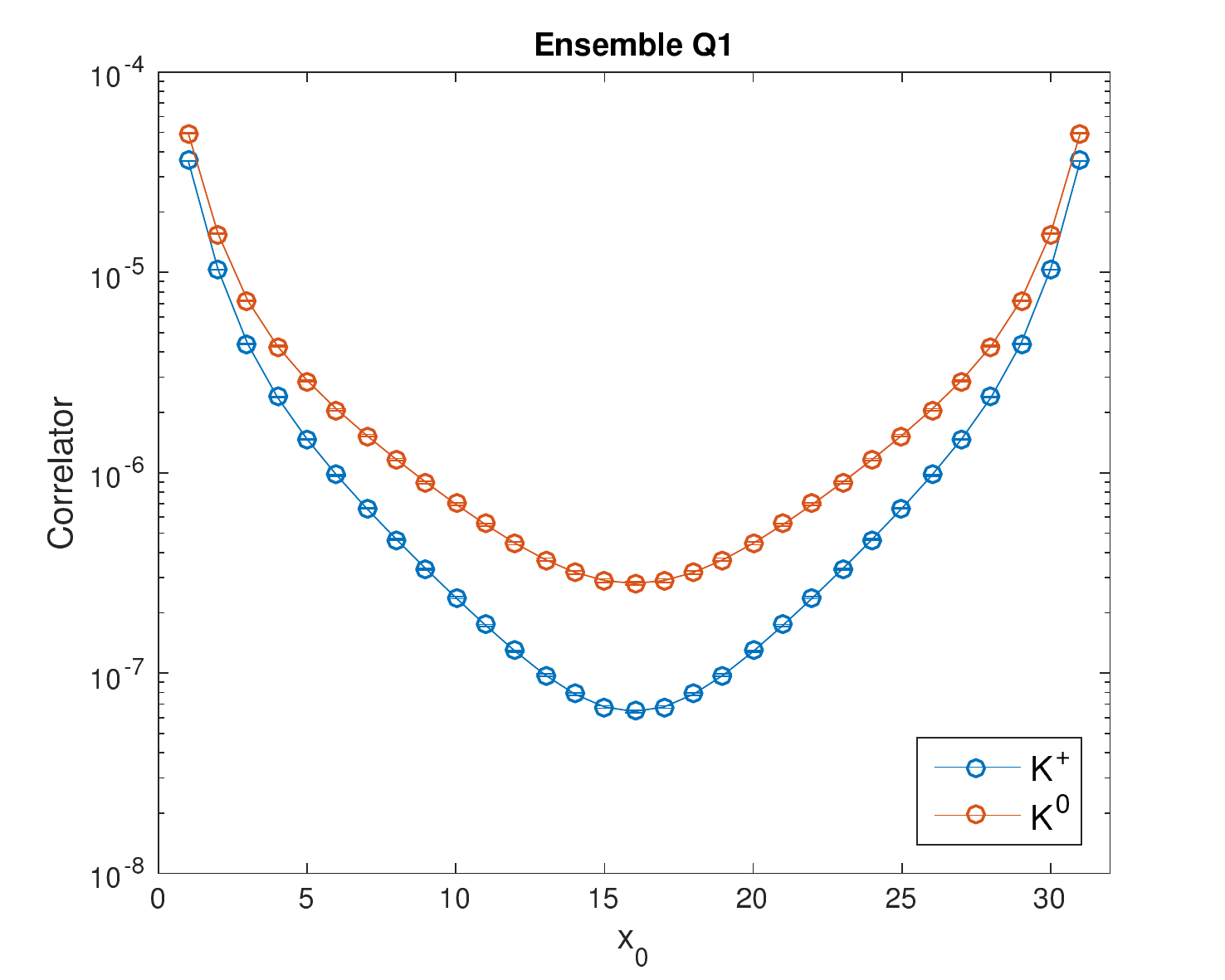}
 \includegraphics[scale=0.45]{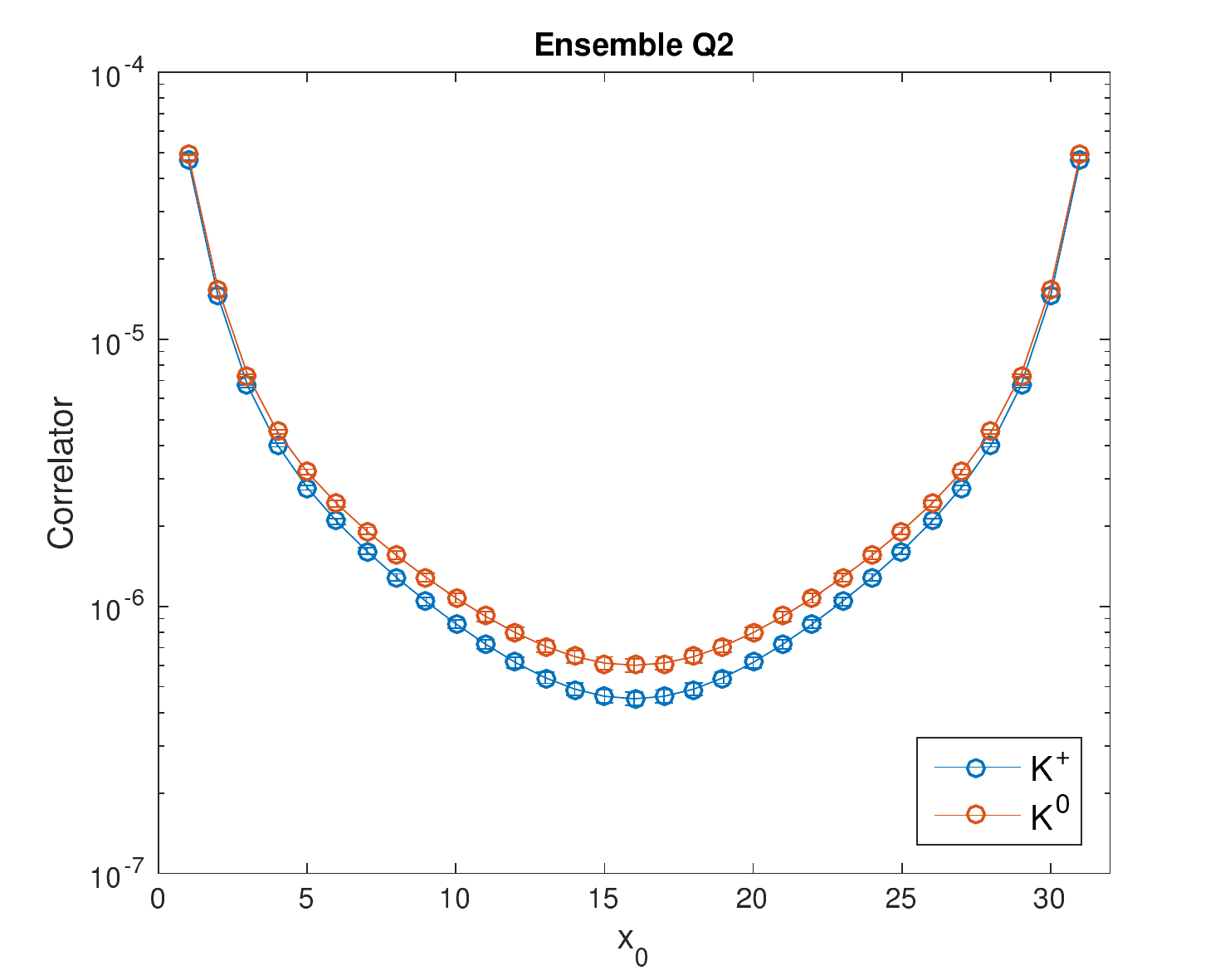} \\
 \includegraphics[scale=0.45]{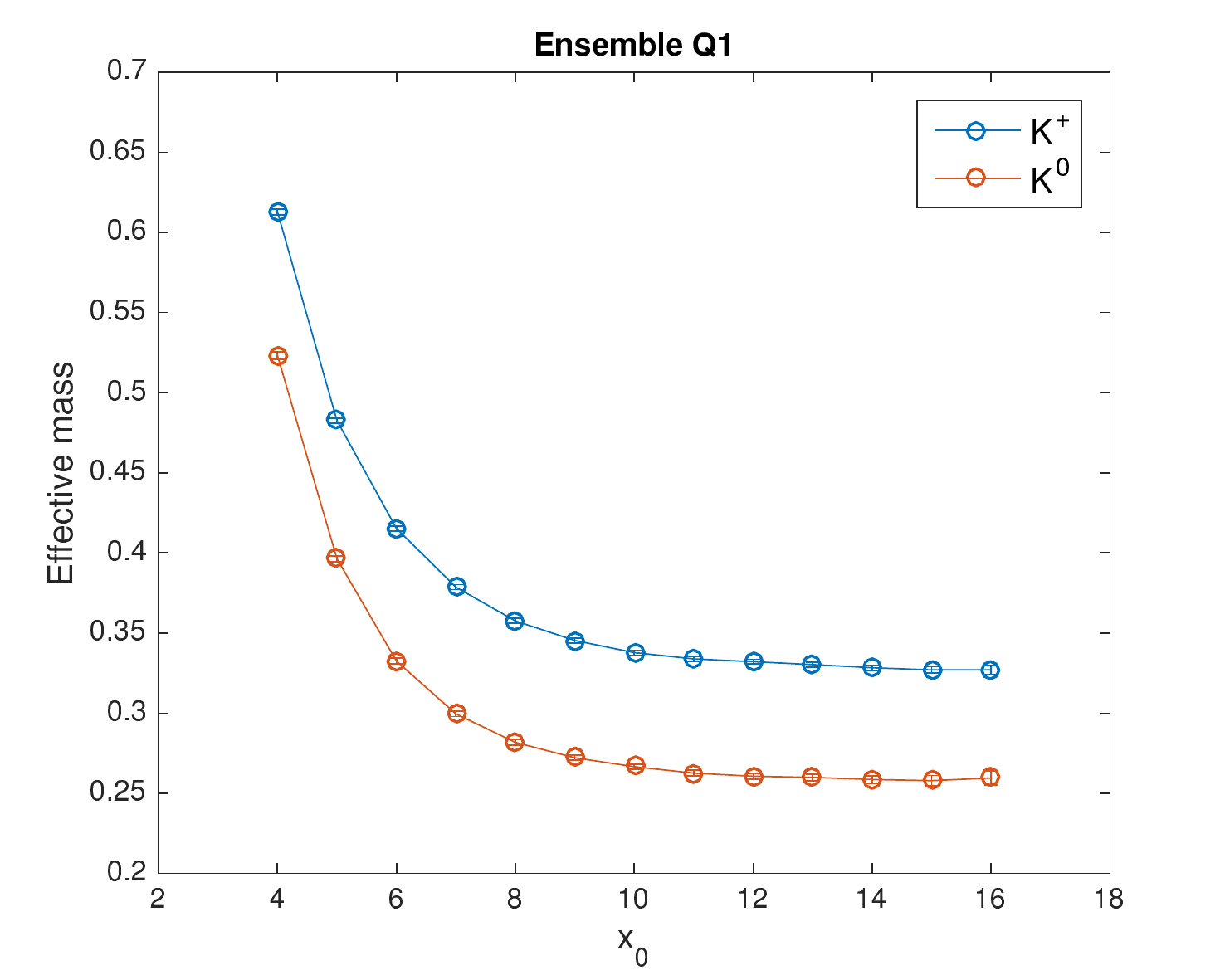}
 \includegraphics[scale=0.45]{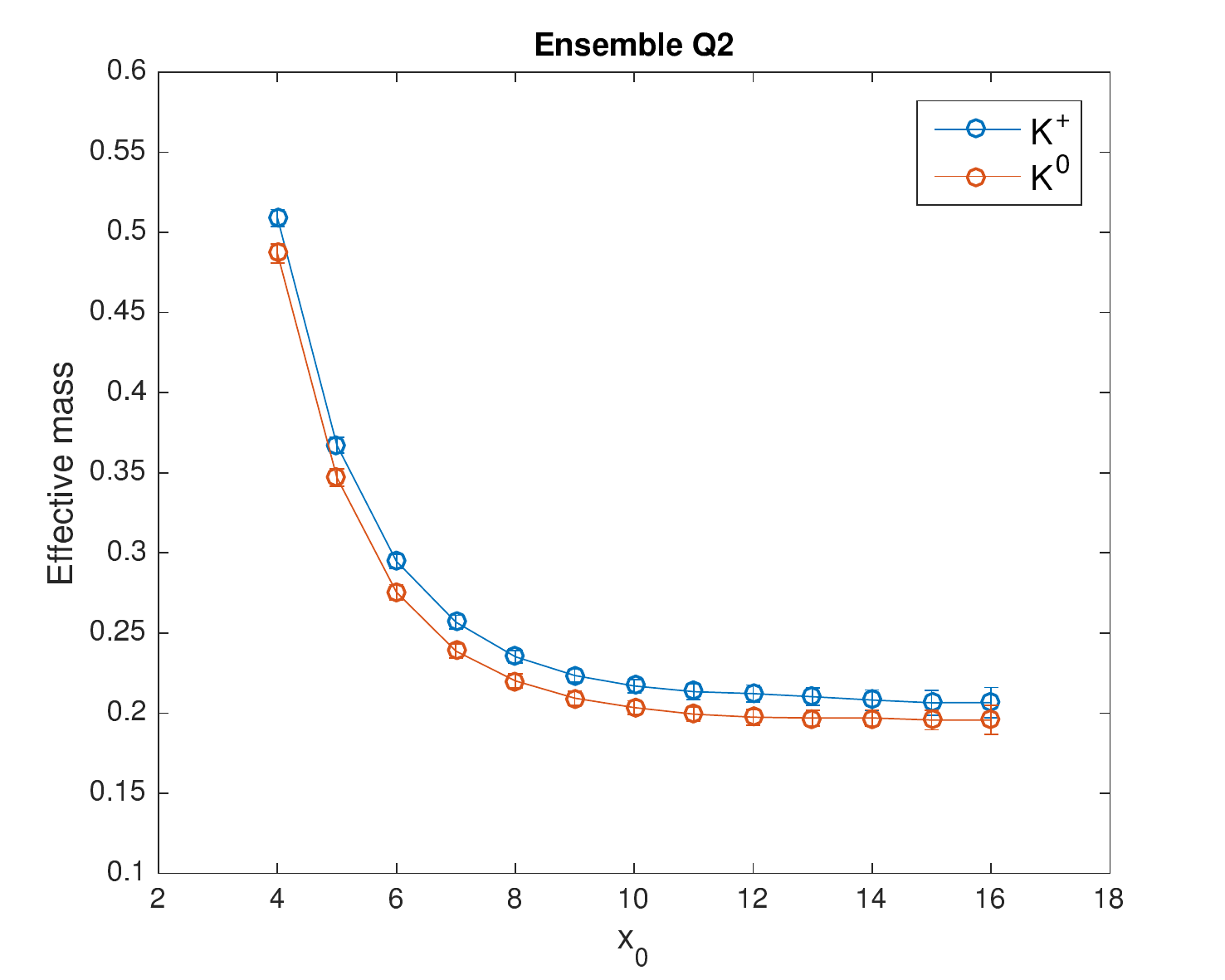} \\
 \includegraphics[scale=0.45]{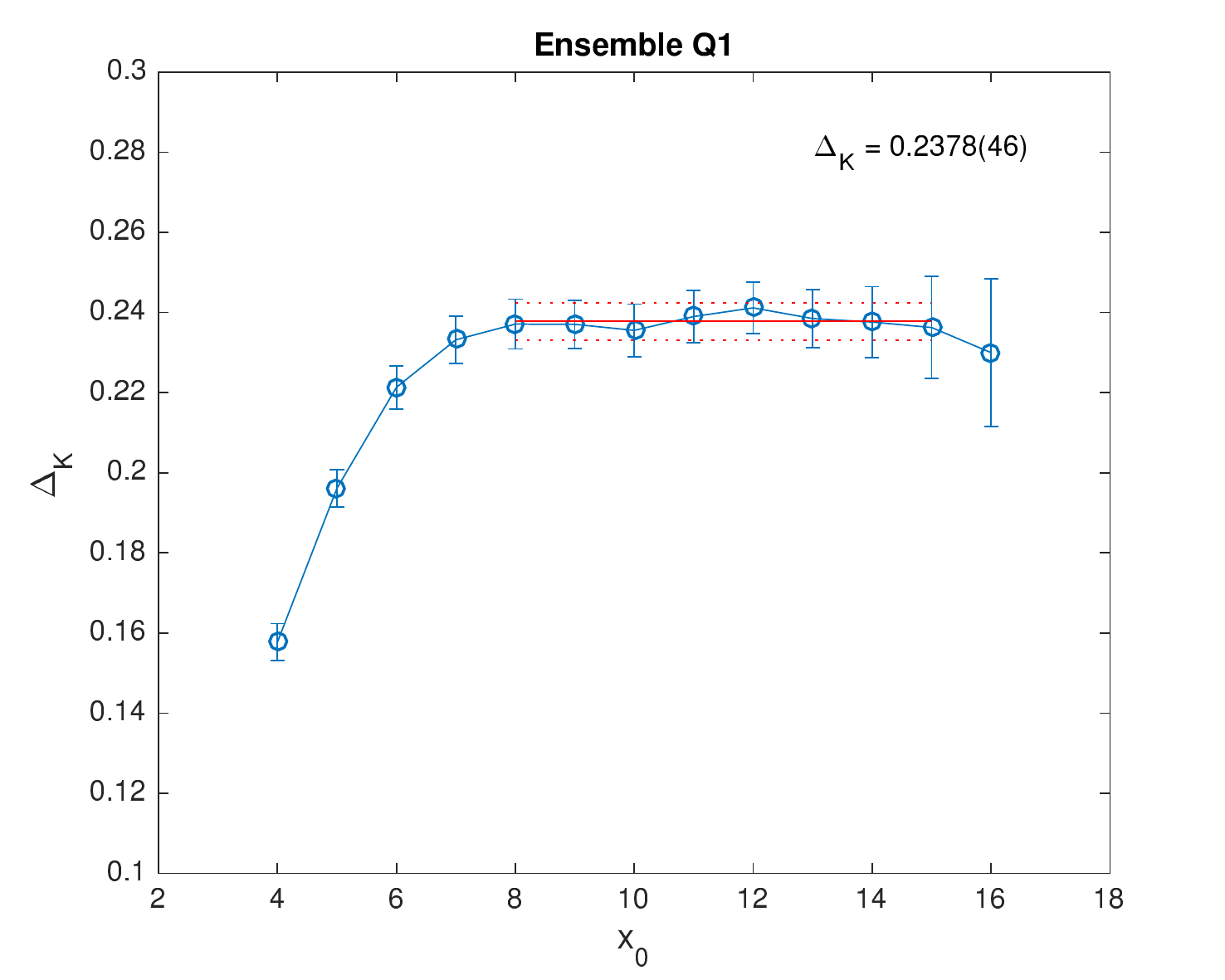}
 \includegraphics[scale=0.45]{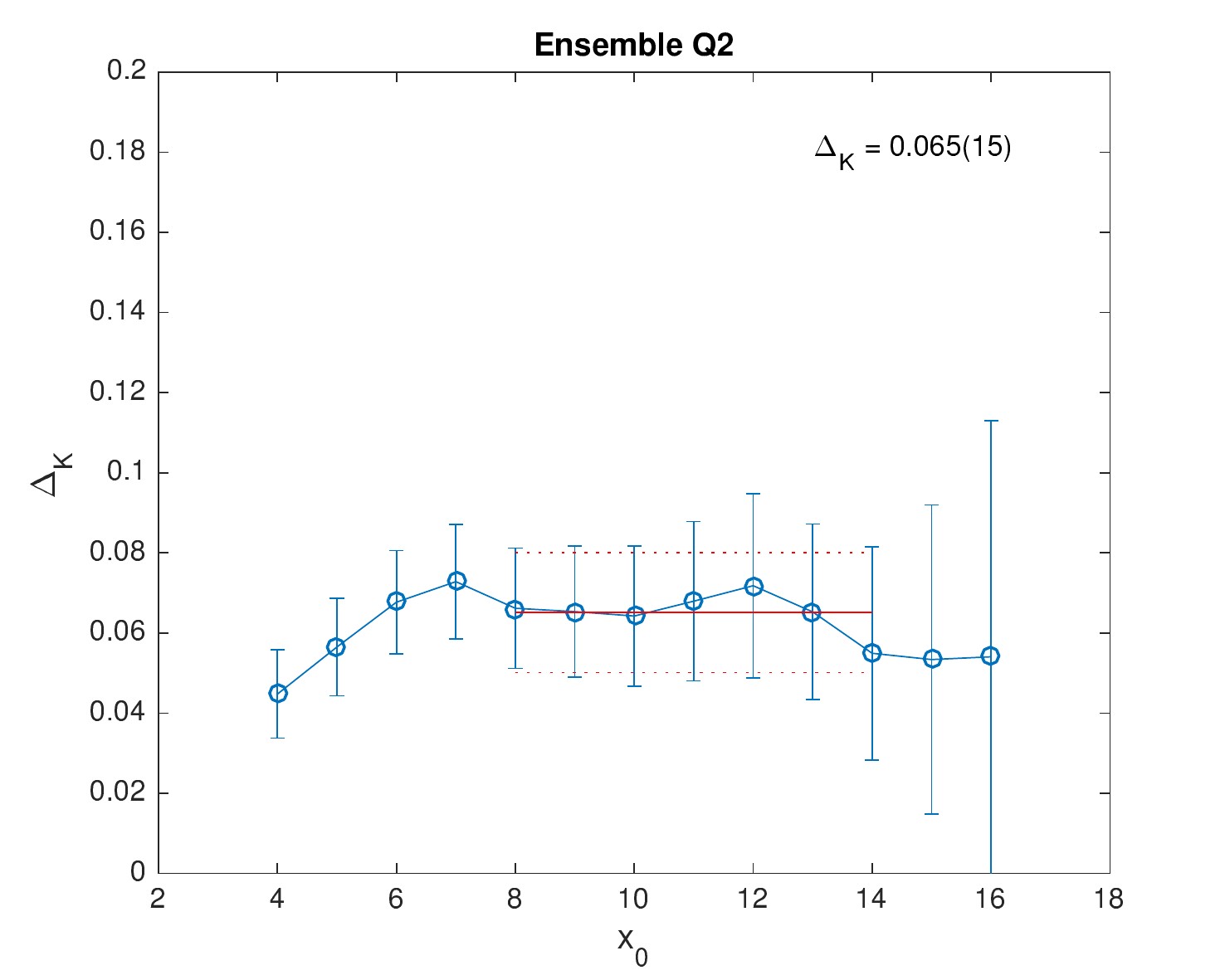}
 \caption{Correlators and effective masses for the charged and neutral kaon state as measured on ensemble Q1 (first column) and Q2 (second column). The relative mass difference $\Delta_K$ is defined in Eq. \eqref{eq:dK}.}
 \label{fig:qcdqed}
\end{figure}

As previously stated, the primary goal of these exploratory QCD+QED simulations is to prove that we can extract the masses of charged hadrons in a gauge invariant way using the operator defined in section~\ref{interp}. In particular we have studied the mass of the charged and neutral kaon. In Fig.~\ref{fig:qcdqed} we show in the first column the results for the Q1 ensemble and in the second column the results for the Q2 ensemble. In the first row we show the correlators for the two states, in the second row the corresponding effective masses, and in the last row we show the effective mass splitting. This effective splitting, denoted $\Delta_K$ in the plots, is defined as the mass difference divided by the average mass, i.e.
\begin{equation}
 \Delta_K = 2\left(\frac{M_{K^+}-M_{K^0}}{M_{K^+}+M_{K^0}}\right)~.
 \label{eq:dK}
\end{equation}
For the Q1 ensemble we observe a large mass splitting with $\Delta_K=0.2378(46)$ due to the unphysically large value of the electromagnetic coupling. As expected, we also observe that the masses increase compared to the QCD simulation (ensemble B1), due to the QED interactions shifting the critical bare masses. On the Q1 ensemble the mass splitting is naturally smaller with $\Delta_K=0.065(15)$. While these simulations are affected by finite volume effects, we are clearly able to extract a statistically significant signal even for the physical value of the electromagnetic coupling constant.

\subsection{Optimizations}
\begin{figure}
 \centering
 \includegraphics[scale=0.60]{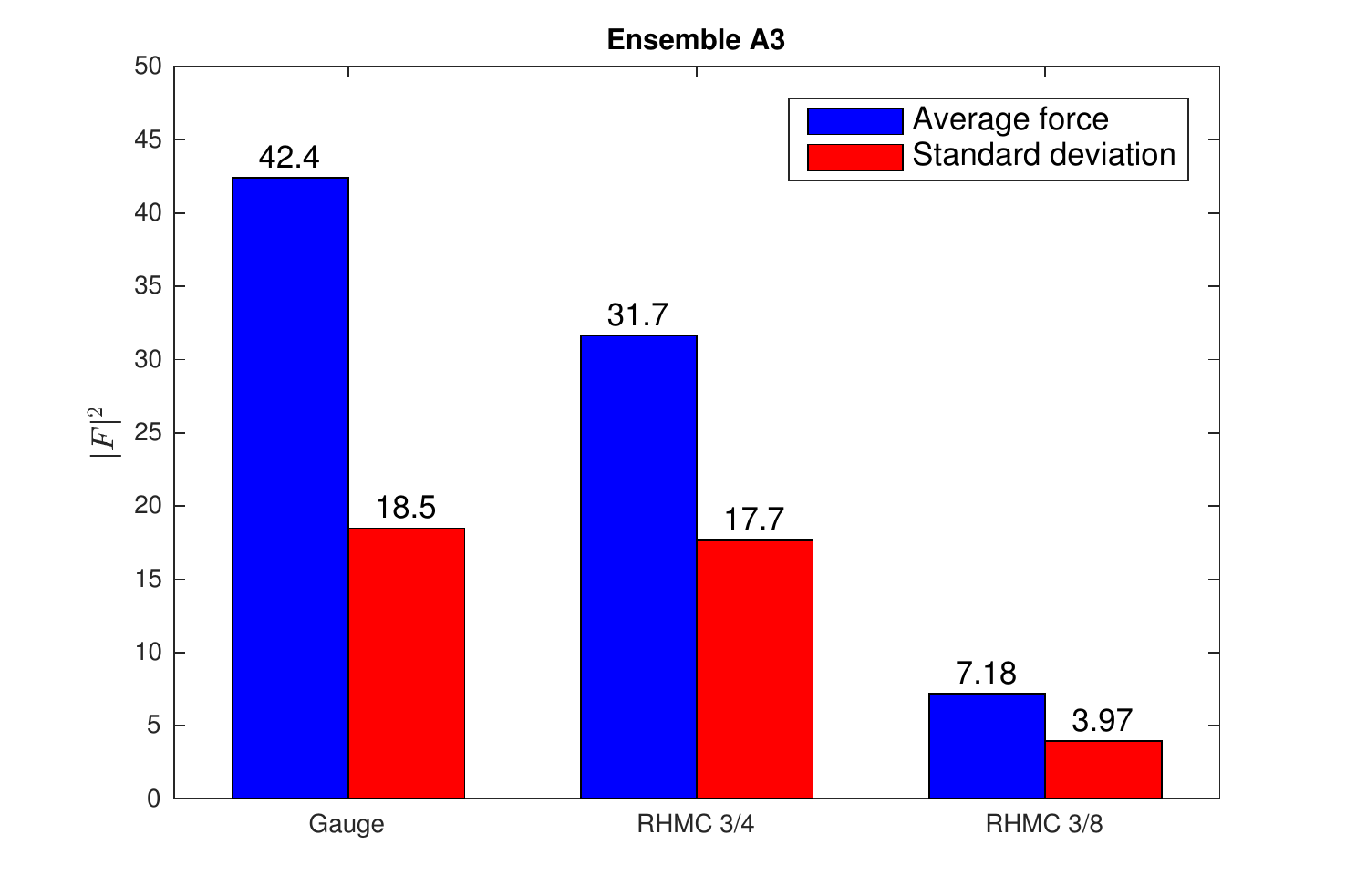}
 \caption{Comparison of the numerical MD forces for the A3 ensemble. The mean value and standard deviation of the fermion force is drastically reduced when decreasing the exponent in the rational approximation.}
 \label{fig:force}
\end{figure}

As argued in section \ref{fermionrep}, the RHMC algorithm is necessary for simulations with \cstar boundary conditions. This will naturally make the simulations significantly more expensive, but there are several ways to reduce the cost, such as pole-splitting for the rational approximation or the introduction of multiple pseudofermions \cite{Clark:2006fx}. We explored the possibility of using multiple pseudofermions for our QCD simulations and we were eventually able to reduce the cost by more than a factor of two\footnote{We use a simple two-level integration scheme with fermions on the outer level and gauge on the inner level.}. In our setup we can simulate $N_f=3$ degenerate fermions with a single pseudofermion when using the fraction $3/4$ in the rational approximation. This was, however, quite expensive because of the relatively ill-conditioned Dirac operator. For this reason we changed the action to include two pseudofermions, each with a fraction of $3/8$, which resulted in a much better conditioned setup. In practice it allowed us to reduce the number of integration steps by roughly a factor of five without loosing acceptance. In Fig.~\ref{fig:force} we show how the mean value and standard deviation of the numerical MD forces decrease when using two pseudofermions for the A3 ensemble, which explains why the number of integration steps can be decreased. On the plot $|F|^2$ is the squared norm of the force vector, and the average is over all positions and directions. Naturally, we also tried using three pseudofermions, but no futher gain was observed in this case.

For our QCD+QED simulations we naturally need two pseudofermions because of the different charge assignments for the quarks, which automatically leads to a reasonably well-conditioned setup. What makes a difference for these simulations is the use of Fourier acceleration for the U(1) field. This was originally implemented to decrease the autocorrelation, but unexpectedly, the number of integration steps could also be decreased by a factor of two, without loosing acceptance, compared to the same simulation without Fourier acceleration. This means that, while the use of Fourier acceleration is slightly more expensive, it is well compensated by the possible decrease in the number of integration steps.

\section{Conclusion}
We presented a modified version of the HiRep code suitable for simulations of QCD and QCD+QED with \cstar boundary conditions. The primary features of the code have been discussed together with some implementation specific details. The code has been used to perform the first exploratory simulations and we have shown that we are able to reproduce known results and that masses of charges hadrons can be extracted in a gauge invariant way, with a signal-to-noise ratio equivalent to that of neutral hadrons. Moreover, we have discussed a few possible ways of optimizing the cost of the simulations.

\section*{Acknowledgements}
The main computational resources were provided by the HPC Wales cluster at Swansea University and the DeIC National HPC Centre at the University of Southern Denmark. M.H. would also like to thank Mauro Papinutto for access to Marconi at CINECA in Italy and Marina Marinkovi\'c for access to Piz Daint at CSCS in Switzerland. M.H. would furthermore like to thank Alberto Ramos and Marina Marinkovi\'c for help and suggestions during the entire extent of the project.

\end{document}